\documentclass[11pt]{article}
\begin{document}
\title
{Note on black hole no hair theorems for massive forms and spin-$\frac12$ fields} 
\author
{Sourav Bhattacharya{\footnote{souravbhatta[AT]hri[DOT]res[DOT]in}}
\\
Harish-Chandra Research Institute, Chhatnag Road, Jhunsi, \\
Allahabad-211019, INDIA.\\
}

\maketitle
\abstract

{We give a proof of the non-perturbative no hair theorems for a massive 2-form field with 3-form field strength for general stationary axisymmetric and static (anti)-de Sitter or asymptotically flat black hole spacetimes with some suitable geometrical properties. The generalization of this result for higher form fields is discussed.
Next, we discuss the perturbative no hair theorems for massive spin-$\frac12$ fields for general static 
backgrounds with electric or magnetic charge.
Some generalization of this result for stationary axisymmetric spacetimes are also discussed. All 
calculations are done in arbitrary spacetime dimensions.}

\hskip 1cm

    {\bf Keywords:} {Stationary axisymmetric black holes, no hair theorems, forms, spin-$\frac12$}

\vskip 1cm
\section{Introduction}
The black hole no hair theorems state that any realistic gravitational collapse must come to a final stationary state 
characterized only by parameters like mass, angular momentum, and charges corresponding to long range gauge fields
(see e.g.~\cite{Chrusciel:1994sn, Heusler:1998ua, Heusler,
Bekenstein:1998aw, Bekenstein:1971hc,Bekenstein:1972ky}, and references therein). The proof
of the no hair theorem for a given matter field for a given black hole spacetime thus essentially involves the 
proof of vanishing of that matter field in the exterior of that spacetime.  

Considerable effort has been given so far to investigate no hair theorems for various matter fields,
such as scalars with or without non-minimal couplings, massive 1-form and spin-2 fields~\cite{Bekenstein:1971hc, Bekenstein:1972ky, Bhattacharya:2007ap,Bhattacharya:2011dq,Bhattacharya:2012yt, Anabalon:2012ta,Anabalon:2009qt,Sen:1998bj}. We refer our reader to~\cite{Anabalon:2012tu, Anabalon:2012sn,Anabalon:2012ih, Acena:2012mr} for some exception to this theorem and to e.g.
\cite{Bhattacharya:2012yt} for a more detailed review on no hair theorems. We 
also refer our reader to~\cite{Johannsen:2012ng, Rodriguez:2011aa,Johannsen:2011dh} for an account of
possible observational consequences related to the black hole no hair theorems.

Since the concern of this paper is to discuss the no hair theorems associated with massive forms and
spin-$\frac12$
fields, let us take a brief account of progress on this topic now. In~\cite{Shiromizu:2011he}, the perturbative
no hair theorems for massless {\it p}-forms with ({\it p}$+$1)-form field strengths
in arbitrary dimensional static spherically symmetric spacetimes were
addressed, by choosing a suitable gauge. Interestingly, when one considers massive 2-form in the context of
a topologically massive gauge theory, a black hole may have a topological charge detectable via Aharonov-Bohm
like effects~\cite{Allen:1989kc}.

The no hair properties for spin-$\frac12$ fields corresponding to various static spherically
symmetric black hole spacetimes has
been discussed in~\cite{Moderski:2008nq, Gibbons:2008gg, Gibbons:2008rs, Nakonieczny:2012df} 
via time dependent perturbation techniques, including the presence of a cosmic string~\cite{Gozdz:2010wv},
which gives topology other than $S^2$.
Interestingly, it was indicated in~\cite{Nakonieczny:2012df}
using the bosonisation scheme techniques that an asymptotically flat black hole spacetime carrying
a non-Abelian charge may face instability if perturbed by a Dirac fermion. A demonstration
of the Price's theorem~\cite{Price:1971gc} for Schwarzschild-de Sitter spacetime for massless spinor
zero modes can be found in~\cite{Chambers:1994sz}.
A proof of non-existence of time-periodic Dirac hair in asymptotically flat static or stationary
axisymmetric spacetimes of dimension four using the variable separated Dirac 
equation~\cite{Chandrasekhar:1985kt} and the properties of the self adjoint Dirac operator
can be found in~\cite{Finster:1998ju,Finster:1998ak, Finster:2001vn, Finster:2000jz, Finster:2005dg, 
Finster:1999tt, Finster:1999ry, Finster:2000vy, Finster:2000ps}. These proofs have also been generalized for the de Sitter black hole spacetime~\cite{Belgiorno:2008mx} (see also \cite{Bhattacharya:2012yt}).

The main concern of this work is higher dimensional general spacetimes satisfying Einstein's 
equations with or without a cosmological constant $\Lambda$, for which 
uniqueness properties including the topology are not yet very well known. 
The uniqueness properties of black hole spacetimes in higher dimensions does 
not seem to have trivial generalizations of what is obtained in dimension four
(like the Birkhoff or Robinson-Carter theorems,~\cite{Chandrasekhar:1985kt}). Consequently,
the statement of such uniquenesses may be quite different and may contain qualitative new features 
in higher spacetime dimensions. Therefore, an essential step in this direction involves the study
of matter fields in such spacetimes, i.e. to check the validity of the no hair theorems. In higher
dimensions, there may exist solutions where variable separation for the equation of motion may be
quite complicated. The situation obviously gets much more involved when we include backreaction
of the matter field. Most importantly, one cannot rule out the existence of more than one
solution with the same geometrical properties.
All these clearly indicate that in higher dimensions we should attempt the problem in a
more unified way, rather than making a case by case study. 
Consequently, proofs of these theorems in such spacetimes should involve a general 
coordinate independent set up, which we describe 
in the next Section for both stationary axisymmetric and
static spacetimes. For static spacetimes we shall not assume
any spatial symmetry (spherical symmetry, for example). We shall not also assume any particular topology. 
The proofs will be coordinate independent and will mainly be based on the symmetry and suitable geometrical
properties of the spacetime and hence matter fields, some reasonable energy conditions and Killing identities.  
We note that the proof of a no hair theorem usually involves the demonstration of
vanishing of a particular matter field by forming vanishing integrals of sum of positive definites, and this chief
characteristic of all such proofs are the same. Though, it may be non-trivial to construct such
integrals depending upon the nature of the matter field and the spacetime, including its dimensionality.
An explicit example of this will be encountered in Section 4, where we shall discuss fermions in charged
black hole backgrounds with arbitrary dimensions. There is a Ricci scalar term in the
squared Dirac equation, and is related to the trace of the energy-momentum tensor of the Maxwell
field. This trace is vanishing in four spacetime dimensions, but not in higher ones. In particular,
we cannot assign a definite sign with it for dyonic black holes. Consequently, we have to manipulate
the calculations by using Killing identity to give it a suitable form.

Based on the set up described in the next section, we give
a proof for no hair theorems for massive 2- and higher forms in stationary axisymmetric
spacetimes (with or without $\Lambda$) with two commuting Killing vector fields and
without ignoring backreaction in Section 3. 
As a corollary, a proof for general static spacetimes is also given. We note that the study of
massive form fields can be particularly interesting and motivating
in the context of dark matters~\cite{Meierovich, Prokopec:2006kr}.
The no hair properties for massive spin-$\frac12$ fields without backreaction will be 
discussed in Section 4. First we shall discuss fermion zero mode solutions in general static
black hole spacetimes with charge. This result is further generalized to general static electrically
charged spacetimes  with $\Lambda\leq0$ for fermions with real frequencies. Next we discuss
the case of stationary axisymmetric (anti)-de Sitter spacetimes with arbitrary number of 
commuting Killing fields.   
The discussions for spin-$\frac12$ fields will be an extension of~\cite{Bhattacharya:2012yt},
where vacuum or $\Lambda$-vacuum stationary axisymmetric spacetimes with two commuting 
Killing fields are addressed. 

We shall use mostly positive signature for the metric $(-,+,+,+,\dots)$, and set $8\pi G=c=\hbar=1$.
\section{The geometrical constructions and assumptions}
Let us start with an outline of the geometry we shall work in 
and derive some useful expressions.
We assume that the spacetime is an {\it n}-dimensional smooth manifold with a Lorentzian metric $g_{ab}$, 
and satisfies Einstein's equations, and there is no naked curvature singularity anywhere in our region
of interest. This means that invariants constructed from the curvature and energy-momentum tensors are
bounded everywhere in our region of interest.
We assume that the spacetime connection is torsion-free.

We assume that any backreacting classical matter energy-momentum tensor satisfies the weak and null
energy conditions, i.e. for any two timelike and null vector fields $t^a$ and $n^a$, we have 
$T_{ab}t^at^b\geq0$ and $T_{ab}n^an^b\geq0$.  
We also assume that for any two future directed timelike vector fields $t_1^a$ and $t_2^a$, the quantity $T_{ab}t_1^at_2^b\geq0$~\cite{Ehlers:2003tv}. This means that the energy density measured by any future directed timelike observer corresponding to any future directed energy current must be positive definite. Interestingly, such energy condition implies that `sufficiently' small bodies in general relativity move along timelike geodesics (see~\cite{Ehlers:2003tv} and references therein). It is also easy to see that this energy condition can in fact be related to the dominant energy condition : for any future directed timelike vector field $t_1^a$, $-T_{ab}t_1^a$ is non-spacelike.

For static spacetimes in arbitrary dimensions, there exists by definition a timelike Killing vector field orthogonal to a family of spacelike hypersurfaces, $\Sigma$. We do not need to assume any spatial symmetry for $\Sigma$.

The case of stationary axisymmetric spacetimes is more complicated which we will describe below. The degree of complications depend on the number of axisymmetric Killing fields non-orthogonal to the timelike Killing field $\xi^a$.    
Let us start with an {\it n}-dimensional stationary axisymmetric spacetime having three commuting Killing fields $(\xi_a,~\phi_a,~\phi^{1}_a)$, respectively generating stationarity and axisymmetries, and non-orthogonal to each other. Any spatial isometry orthogonal to $\xi^a$ may be present, but will not complicate the calculations. The generalization to higher number of non-orthogonal Killing fields will be clear from the following discussions. It will be a generalization of
\cite{Bhattacharya:2013trx} (see also references therein), in which 
details for such spacetimes with two commuting Killing fields can be found.

We have
\begin{eqnarray}
\nabla_{(a}\xi_{b)} = 0,\quad\nabla_{(a}\phi_{b)}=0,\quad\nabla_{(a}\phi^1_{b)}=0,\nonumber\\
\pounds_{\xi}\phi^b=0=\pounds_{\xi}\phi^{1b},\quad \pounds_{\phi}\phi^{1b}=0.
\label{g1}
\end{eqnarray}
We assume that the ({\it n}$-$3)-dimensional
spacelike surfaces orthogonal to $(\xi_a,~\phi_a,~\phi^{(1)}_a)$ form integral submanifolds, which implies~\cite{Wald:1984rg},
\begin{eqnarray}
\phi_{[a}\phi^{1}_b\xi_c\nabla_d \xi_{e]}=\phi_{[a}\xi_b\phi^{1}_c\nabla_d \phi^1_{e]}=\phi^1_{[a}\xi_b\phi_c\nabla_d \phi_{e]}=0.
\label{g9}
\end{eqnarray}
For convenience, we construct a set of basis vectors $(\chi_a,~\phi_a,~\widetilde{\phi}_a)$ as
\begin{eqnarray}
\chi_a=\xi_a+\alpha_1 \phi_a+\alpha_2\phi^1_a,~\widetilde{\phi}_a=\phi^1_a+\lambda \phi_a,
\label{g10}
\end{eqnarray}
so that $\chi^a$, $\phi^a$ and $\widetilde{\phi}_a$ are orthogonal to each other everywhere. This requirement fixes the functions $\alpha_i(x)$ and $\lambda(x)$. Let the norms of $(\xi^a,~\phi^a,~\widetilde{\phi}^a)$ be 
$(-\lambda^{\prime 2},~f^2,~\widetilde{f}^2)$ respectively.
Using the second of the above equations into the first, we rewrite the basis as
\begin{eqnarray}
\chi_a=\xi_a+\alpha \phi_a+\widetilde{\alpha} \widetilde{\phi}_a,~\widetilde{\phi}_a=\phi^1_a+\lambda \phi_a,
\label{g11}
\end{eqnarray}
so that $\alpha=-\frac{\xi\cdot\phi}{f^2}$, $ \widetilde{\alpha}=-\frac{\xi\cdot \widetilde{\phi}}{\widetilde{f}^2}$,
and $\lambda=-\frac{\phi^1\cdot\phi}{f^2}$. We find 
\begin{eqnarray}
\chi_a\chi^a=-\beta^2=-\left(\lambda^{\prime 2}+\alpha^2f^2+\widetilde{\alpha}^2\widetilde{f}^2\right),
\label{g12}
\end{eqnarray}
so that $\chi^a$ is timelike when $\beta^2\geq 0$. The price we have paid doing this orthogonalization is that, $\widetilde{\phi}_a$ and $\chi_a$ are not Killing fields,
\begin{eqnarray}
\nabla_{(a}\widetilde{\phi}_{b)}=\phi_{(a}\nabla_{b)}\lambda,\quad
\nabla_{(a}\chi_{b)}=\widetilde{\phi}_{(a}\nabla_{b)}\widetilde{\alpha}+\phi_{(a}\left(\nabla_{b)}\alpha+\widetilde{\alpha}\nabla_{b)} \lambda\right).
\label{g13}
\end{eqnarray}
It is easy to see using the commutativity of the Killing fields that
\begin{eqnarray}
\pounds_{\chi}\beta=\pounds_{\chi}f=\pounds_{\chi}\widetilde{f}=\pounds_{\chi}\alpha=\pounds_{\chi}\widetilde{\alpha}=\pounds_{\chi}\lambda=0, \nonumber\\
\pounds_{\phi}\beta=\pounds_{\phi}f=\pounds_{\phi}\widetilde{f}=\pounds_{\phi}\alpha=\pounds_{\phi}\widetilde{\alpha}=\pounds_{\phi}\lambda=0,\nonumber\\
\pounds_{\widetilde{\phi}}\beta=\pounds_{\widetilde{\phi}}f=\pounds_{\widetilde{\phi}}\widetilde{f}=\pounds_{\widetilde{\phi}}\alpha=\pounds_{\widetilde{\phi}}\widetilde{\alpha}=\pounds_{\widetilde{\phi}}\lambda=0,
\label{g13'}
\end{eqnarray}
and
\begin{eqnarray}
\pounds_{\phi} \widetilde{\phi^a}=0=\pounds_{\phi}\widetilde{\phi_a}, \quad
\pounds_{\phi}\chi^a=0=\pounds_{\phi}\chi_a,~\pounds_{\widetilde{\phi}}\chi^a=0=\pounds_{\widetilde{\phi}}\chi_a.
\label{g13''}
\end{eqnarray}
In terms of our new basis the integrability conditions (\ref{g9}) become
\begin{eqnarray}
\phi_{[a}\widetilde{\phi}_b\chi_c\nabla_d \chi_{e]}=\phi_{[a}\chi_b\widetilde{\phi}_c\nabla_d \widetilde{\phi}_{e]}=\widetilde{\phi}_{[a}\chi_b\phi_c\nabla_d \phi_{e]}=0,
\label{g14}
\end{eqnarray}
which permit solutions of the form
\begin{eqnarray}
\nabla_{[a} \chi_{b]}=\mu_{1[a}\chi_{b]}+\mu_{2[a}\phi_{b]}+\mu_{3[a}\widetilde{\phi}_{b]}+\nu_1\chi_{[a}\phi_{b]}
+\nu_2\chi_{[a}\widetilde{\phi}_{b]}+\nu_3\phi_{[a}\widetilde{\phi}_{b]},\nonumber\\
\nabla_{[a} \phi_{b]}=\mu_{4[a}\chi_{b]}+\nu_{5[a}\phi_{b]}+\mu_{6[a}\widetilde{\phi}_{b]}+\nu_4\chi_{[a}\phi_{b]}
+\nu_5\chi_{[a}\widetilde{\phi}_{b]}+\nu_6\phi_{[a}\widetilde{\phi}_{b]},\nonumber\\
\nabla_{[a} \widetilde{\phi}_{b]}=\mu_{7[a}\chi_{b]}+\mu_{8[a}\phi_{b]}+\mu_{9[a}\widetilde{\phi}_{b]}+\nu_7\chi_{[a}\phi_{b]}
+\nu_8\chi_{[a}\widetilde{\phi}_{b]}+\nu_9\phi_{[a}\widetilde{\phi}_{b]},
\label{g15}
\end{eqnarray}
where $\mu_{ia}$ are 1-forms orthogonal to $\chi^a$, $\phi^a$ and $\widetilde{\phi}^a$ and $\nu_i(x)$ are functions which we have to determine for our purpose. 

Contracting the first of Eq.s~(\ref{g15}) by $\chi^a\phi^b$, using Eq.s~(\ref{g13'}), the orthogonality of $\chi^a$, $\phi^a$, and $\widetilde{\phi}^a$, keeping in mind that $\mu_{ia}$'s are orthogonal to ($\chi^a,~\phi^a,~\widetilde{\phi}^a$), and the Killing equation for $\phi_a$ gives $\nu_1(x)=0$. Similarly, contraction with 
$\chi^a\widetilde{\phi}^b$ and $\phi^a\widetilde{\phi}^b$ and use of Eq.s~(\ref{g13}), (\ref{g13'}), (\ref{g13''}) and the orthogonalities give $\nu_2(x)=0=\nu_3(x)$ respectively. Similarly we find that all the other $\nu_i(x)$'s vanish identically.

Let us now determine the 1-forms $\mu_{ia}$. Contracting the first of Eq.s~(\ref{g15}) by $\chi^a$, using Eq.s~(\ref{g13}), (\ref{g13'}), and the orthogonality between $\chi^a$, $\phi^a$, $\widetilde{\phi}^a$, we find $\mu_{1b}=2\beta^{-1}\nabla_b \beta$. Contracting the equation with $\phi^b$ gives
\begin{eqnarray}
\mu_{2a}=f^{-2}\phi^b\left(\nabla_a\chi_b-\nabla_b\chi_a\right)=-f^{-2}\pounds_{\phi}\chi_a=0,
\label{g16'}
\end{eqnarray}
by the second of Eq.s~(\ref{g13''}). Next we contract the equation with $\widetilde{\phi}^b$ to find    
\begin{eqnarray}
\mu_{3a}=\widetilde{f}^{-2}\widetilde{\phi}^b\left(\nabla_a\chi_b-\nabla_b\chi_a\right)=-\widetilde{f}^{-2}\pounds_{\widetilde{\phi}}\chi_a=0,
\label{g16}
\end{eqnarray}
by the last of Eq.s~(\ref{g13''}). Putting these all in together we find
\begin{eqnarray}
\nabla_{[a}\chi_{b]}=2\beta^{-1}\left(\chi_b\nabla_a\beta-\chi_a\nabla_b\beta\right),
\label{g17}
\end{eqnarray}
which implies $\chi_{[a}\nabla_{b}\chi_{c]}=0$, and hence $\chi_a$ is orthogonal to the family of
({\it n}$-$1)-dimensional spacelike hypersurfaces, say $\Sigma$, which contain $\phi_a$ and $\widetilde{\phi}_a$. Eq.~(\ref{g17}) and the last of Eq.s~(\ref{g13}) give an useful expression,
\begin{eqnarray}
\nabla_{a}\chi_{b}=\beta^{-1}\left(\chi_b\nabla_a\beta-\chi_a\nabla_b\beta\right)+\frac12\widetilde{\phi}_{(a}\nabla_{b)}\widetilde{\alpha}+\frac12\phi_{(a}\left(\nabla_{b)}\alpha+\widetilde{\alpha}\nabla_{b)} \lambda\right).
\label{g18}
\end{eqnarray}
Similarly we can solve for $\nabla_a\phi_b$ and $\nabla_a\widetilde{\phi}_b$, but we do not need their explicit expressions for our present purpose. 
%
%
%
%
Let us consider a 1-form $\mu_a$ on $\Sigma$,    
\begin{eqnarray}
\mu_a:=\nabla_a\beta^2.
\label{g21}
\end{eqnarray}
On any $\beta^2=0$ hypersurface ${\cal{H}}$
\begin{eqnarray}
\nabla_a\beta^2=2\kappa\chi_a,
\label{g22}
\end{eqnarray}
where $\kappa$ is a function on ${\cal{H}}$. The above equation follows from Eq.~(\ref{g17}), 
\begin{eqnarray}
\chi_{[b}\nabla_{a]}\beta^2\vert_{\beta^2\to 0}=\beta^2\partial_{[a}\chi_{b]}\vert_{\beta^2\to 0}\to 0.
\label{g22'}
\end{eqnarray}
It is clear from Eq.~(\ref{g21}) that $\mu_a$ coincides with $\chi_a$ and becomes null on ${\cal{H}}$. It is easy to see using the torsion-free condition that $\pounds_{\chi}\kappa=0$.

A Killing or true horizon of this spacetime is any $\beta^2=0$ null hypersurface ${\cal{H}}$. This requires a proof, which is the following.

Let us write $\chi_a$ in terms of the Killing fields, $\chi_a=\xi_a+\alpha_1 \phi_a+\alpha_2\phi^1_a$ (Eq.~(\ref{g10})).
Let $\tau$ be the parameter along $\chi_a$, i.e. $\chi^a\nabla_a\tau:=1$. Let $c$ be a constant along $\chi_a$ and we define a 1-form
$k_a=e^{-c\tau}\chi_a$. We compute using $\chi_a\phi^a=0=\chi^a\phi^1_a$, the fact that $\pounds_{\chi}\alpha_1=0=\pounds_{\chi}\alpha_2$ (follow from the commutativity of Killig fields), 
\begin{eqnarray}
k^a\nabla_ak_b=\frac12e^{-2c\tau}\left[\nabla_b\beta^2-2c\chi_b\right].
\label{g23}
\end{eqnarray}
We further compute
\begin{eqnarray}
k_a\nabla_bk_c-k_b\nabla_ak_c=e^{-2c\tau}\left[\chi_a\nabla_b\chi_c-\chi_b\nabla_a\chi_c+c\chi_b\chi_c\nabla_a\tau-c\chi_a\chi_c\nabla_b\tau\right].
\label{g24}
\end{eqnarray}
Now let $\hat{h}^{ab}$ be the induced metric on the (n-2)-dimensional hypersurface orthogonal to both $\mu^a$ and $\chi^a$,
\begin{eqnarray}
\hat{h}^{ab}=f^{-2}\phi^a\phi^b+f_{1}^{-2}\phi^{1a}\phi^{1b}+f_{12}^{-2}\left(\phi^a\phi^{1b}+\phi^b\phi^{1a}\right)+\hat{h'}^{ab},
\label{g25}
\end{eqnarray}
where $f_{1}^{-2}$ is the norm of $\phi^{1a}$, and $ f_{12}^2=\phi_a\phi^{1a}$, and $\hat{h'}^{ab}$ is the induced metric on the remaining (n-4)-dimensional spacelike surfaces, orthogonal to both $\phi_a$ and $\phi^{1a}$. We contract Eq.~(\ref{g24}) with $\hat{h}^{bc}$,
\begin{eqnarray}
k_a\hat{h}^{bc}\nabla_bk_c=\frac12k_ae^{-c\tau}\hat{h}^{bc}\nabla_{(b}\chi_{c)}.
\label{g26}
\end{eqnarray}
Using $\nabla_{(b}\chi_{c)}= \phi_{(b}\nabla_{c)}\alpha_1+\phi^1_{(b}\nabla_{c)}\alpha_2$, and the fact that $\pounds_{\phi}\alpha_{(1,2)}=0=\pounds_{\phi^1}\alpha_{(1,2)}$, we get
\begin{eqnarray}
\hat{h}^{bc}\nabla_bk_c=0.
\label{g27}
\end{eqnarray}
Next we contract Eq.~(\ref{g24}) with the combination : \\$\zeta^{[bc]}=\left(\phi^{[b}\phi^{1c]}+\sum_{i=1}^{n-4}\phi^{[b}X_i^{c]}+\sum_{i=1}^{n-4}\phi^{[1b}X_i^{c]}+\sum_{i,j=1,i\neq j}^{n-4}X^{[b}_iX^{c]}_j\right)$, where $X^a_i\vert_{i=1}^{n-4}$ are basis vectors of $\hat{h'}^{ab}$ in Eq.~(\ref{g25}), we find using Eq.~(\ref{g18}) 
\begin{eqnarray}
\zeta^{[bc]}\nabla_bk_c=0.
\label{g28}
\end{eqnarray}
Next we contract Eq.~(\ref{g24}) with $\zeta^{(bc)}=\left(\phi^{(b}\phi^{1c)}+\sum_{i=1}^{n-4}\phi^{(b}X_i^{c)}+\sum_{i=1}^{n-4}\phi^{(1b}X_i^{c)}+\sum_{i,j=1}^{n-4}X^{(b}_iX^{c)}_j\right)$ to find
\begin{eqnarray}
\zeta^{(bc)}\nabla_bk_c=\frac12e^{-c\tau}\zeta^{(bc)}\nabla_{(b}\chi_{c)}=\frac12e^{-c\tau}\left(\sum_{i=1}^{n-4}\phi^{(b}X_i^{c)}+\sum_{i=1}^{n-4}\phi^{(1b}X_i^{c)}\right)\left( \phi_{(b}\nabla_{c)}\alpha_1+\phi^1_{(b}\nabla_{c)}\alpha_2  \right).
\label{g29}
\end{eqnarray}
Let us now consider the $\beta^2=0$ surface ${\cal{H}}$. Following~\cite{Wald:1984rg}, we shall now construct a null geodesic congruence on ${\cal{H}}$. If we choose $c=\kappa$ on ${\cal{H}}$, Eq.s~(\ref{g22}), (\ref{g23}) show that the vector field $k^a$ is a null geodesic on ${\cal{H}}$. The Raychaudhuri equation for the null geodesic congruence $k^a$ reads~\cite{Wald:1984rg}
\begin{eqnarray}
\frac{d\theta}{ds}=-\frac{1}{\left(n-2\right)}\theta^2-\sigma_{ab}\sigma^{ab}+\omega_{ab}\omega^{ab}-R_{ab}k^ak^b,
\label{g30}
\end{eqnarray}
where $s$ is an affine parameter, and $\theta$, $\sigma_{ab}$ and $\omega_{ab}$ are respectively the expansion, shear and rotation of the congruence given by
\begin{eqnarray}
\theta=\hat{h}^{ab}\nabla_ak_b,\quad \sigma_{ab}=\nabla_{(a}k_{b)}-\frac{1}{(n-2)}\theta \hat{h}_{ab}, \quad \omega_{ab}=\nabla_{[a}k_{b]},
\label{g31}
\end{eqnarray}
where all the derivatives are taken on the spacelike ({\it n}$-$2)-plane orthogonal to $\chi^a$ or $\mu^a$
on ${\cal{H}}$. Eq.s~(\ref{g27}), (\ref{g28}) show $\theta=0=\omega_{ab}$ on ${\cal{H}}$ for the null geodesic congruence $k^a$. Then using Eq.s~(\ref{g29}), (\ref{g31}), the Einstein equations $R_{ab}-\frac{1}{2-n}\left[T-2\Lambda\right]g_{ab}=T_{ab}$ into Eq.~(\ref{g30}), we find
\begin{eqnarray}
 e^{-2\kappa \tau}\left(\phi_a\nabla_b\alpha_1+ \phi^1_a\nabla_b\alpha_2\right)\left(\phi^a\nabla^b\alpha_1+ \phi^{1a}\nabla^b\alpha_2\right)= -2T_{ab}k^ak^b\leq 0,
\label{g32}
\end{eqnarray}
since we have assumed that any backreacting matter energy-momentum tensor satisfies the null energy condition. The left hand side is a spacelike inner product and hence must be positive definite. Therefore the left hand side must vanish on ${\cal{H}}$ to avoid any contradiction. We also note that on ${\cal{H}}$, $\chi_a$ coincides with $\nabla_a\beta^2$, and $\pounds_{\chi}\alpha_1=0=\pounds_{\chi}\alpha_2$. All these suggest that 
$\alpha_1$ and $\alpha_2$ are constants on any $\beta^2=0$ hypersurface, so that $\chi_a$ becomes a null Killing field there and hence any null hypersurface ${\cal{H}}$ is a Killing horizon of the stationary axisymmetric geometry we are considering. 

Then following similar steps as in four spacetime dimensions~\cite{Wald:1984rg}, we can show that $\kappa$ is a constant 
on ${\cal{H}}$. 

For spin-$\frac12$ fields we shall take $\mu_a=\nabla_a\beta^2$ to be one of the basis vectors on $\Sigma$.
It is clear from this choice~(Eq.s~(\ref{g21}),~(\ref{g22})) that our calculations for such fields will be valid for non-extremal or near-extremal solutions ($\kappa\neq0$), but not for the strictly extremal $\kappa=0$ case. 

The projector $h_{a}{}^b$ which projects tensors onto the spacelike hypersurfaces $\Sigma$ is given by
\begin{eqnarray}
h_{a}{}^{b}=\delta_{a}{}^{b}+\beta^{-2}\chi_a\chi^b.
\label{g6}
\end{eqnarray}
Let $D_a$ be the spacelike induced derivative : $D_a\equiv h_{a}{}^{b}\nabla_b$. We have~\cite{Wald:1984rg}
\begin{eqnarray}
D_aT_{a_1a_2\dots}{}^{b_1b_2\dots}:=h_{a}{}^{b}
 h_{a_1}{}^{c_1}\dots h^{b_1}{}_{d_1}\dots\nabla_b
T_{c_1c_2\dots}{}^{d_1d_2\dots},
\label{g7}
\end{eqnarray}
where $T$ is tangent to $\Sigma$, $T_{a_1a_2\cdots}{}^{b_1b_2\cdots} :=
h_{a_1}{}^{c_1}\cdots h^{b_1}{}_{d_1}\cdots T_{c_1c_2\cdots}{}^{d_1d_2\cdots}$. 

Now it is clear that we can generalize the above calculations by adding more commuting Killing fields non-orthogonal to $\xi^a$. For example, for four commuting non-orthogonal Killing fields $(\xi^a,\phi^a,\phi^{1a},\phi^{2a})$, we will
have $\chi_a=\xi_a+\alpha\phi_a+\alpha_2\phi^{1}_{a}+\alpha_3\phi^{2}_{a}$. Next we can orthogonalize the axisymmetric
Killing fields to write the analogous form of Eq.s~(\ref{g11}). The integrability conditions (\ref{g9}) or
(\ref{g14}) now involves four vector fields and we can solve them as earlier. Thus the process goes on for higher 
number of Killing fields.

For two commuting Killing vector fields $\xi^a$, and $\phi^a$, we have $\chi_a=\xi_a+\alpha \phi_a$, with $\alpha=-\frac{\xi\cdot \phi}{\phi\cdot \phi}$. Eq.~(\ref{g18}) in this case becomes~\cite{Bhattacharya:2013trx}
\begin{eqnarray}
\nabla_a\chi_{b}=\beta^{-1}\chi_{[b}\nabla_{a]}\beta +
\frac12\phi_{(a}\nabla_{b)}\alpha.  
\label{g5'n}
\end{eqnarray}
We shall also require the following expression for two commuting Killing fields:
\begin{eqnarray}
\nabla_a\phi_{b}=f^{-1}\phi_{[b}\nabla_{a]}f+\frac{f^2}{2\beta^2}\chi_{[a}\nabla_{b]}\alpha.
\label{g9''}
\end{eqnarray}
We shall also require the projector in this case onto the integral ({\it n}$-$2)-planes (say $\overline{\Sigma}$) orthogonal to both $\chi^a$, $\phi^a$ 
\begin{eqnarray}
\Pi_{a}{}^{b}=\delta_{a}{}^{b}+\beta^{-2}\chi_a\chi^b-f^{-2}\phi_a\phi^b. 
\label{g8}
\end{eqnarray}
We shall denote the induced connection on $\overline{\Sigma}$ by $\overline{D}$, defined similarly as what we did for $\Sigma$

For the cosmological constant to be vanishing or negative, we assume the spacetime to be respectively asymptotically flat
or anti-de Sitter. For $\Lambda>0$, we shall assume the existence of a de Sitter Killing
horizon (with $\beta^2=0$) surrounding the black hole horizon. Apart from the existence of the cosmological horizon as an outer
boundary and regularity, no precise asymptotics on spacetime or matter fields will be imposed for the de Sitter case. 

We assume that any physical matter field, or any observable concerning the matter field also obeys the symmetries of the
spacetime, be it continuous or discrete~\cite{Bekenstein:1971hc, Bekenstein:1972ky, weinberg}. Thus if
$X$ is a physical matter field or a component of it, or an observable quantity associated with it, we must have its Lie derivative vanishing along a Killing field. Likewise, if the spacetime has any discrete symmetry, we shall assume any 
physical matter field obeys the symmetry. For static spacetimes we have a time reversal symmetry $\xi^a\to
-\xi^a$, whereas for stationary axisymmetric spacetimes with two commuting Killing fields have symmetry under the
simultaneous reflections $\xi^a\to-\xi^a $ and $\phi^a\to -\phi^a$. 

As we have seen above that the classical energy conditions play crucial role in constructing the geometry, unlike form fields, we shall ignore backreaction of the spinors on the spacetime since spinors do not obey any classical energy condition~\cite{Chandrasekhar:1985kt, Penrose:1985jw}. We shall also assume for the spin-$\frac12$ case following~\cite{Bekenstein:1971hc, Bekenstein:1972ky}
  that the Compton wavelength of the massive field is much small compared to the length scale of the black hole horizon.
We note that this is not a strong assumption, since if we have a spinor having Compton wavelength comparable to the black hole horizon size, the assumption of negligible backreaction may be invalidated.
 
This completes the necessary geometrical set up and clarifies all assumptions, and we shall now go into the proofs.

\section{Massive forms}
We shall start with a free theory of massive 2-form field $B_{ab}$ with 3-form field strength $H_{abc}$,
\begin{eqnarray}
{\cal{L}}&=&-\frac{1}{12}H_{abc}H^{abc}-\frac{m^2}{4}B_{ab}B^{ab},\nonumber\\
H_{abc}&=&\nabla_{a}B_{bc}+\nabla_{b}B_{ca}+\nabla_{c}B_{ab}.
\label{b1}
\end{eqnarray}
The equation of motion for the $B$ field reads
\begin{eqnarray}
\nabla_aH^{abc}-m^2B^{bc}=0.
\label{b2}
\end{eqnarray}
We shall consider this theory in a stationary axisymmetric spacetime with two commuting non-orthogonal Killing
fields $\xi^a$ and $\phi^a$. An explicit example with $\Lambda=0$ of such an {\it n}-dimensional spacetime can be found 
in~\cite{perry}.

We have by symmetry requirement 
\begin{eqnarray}
\pounds_{\xi}B_{ab}=0=\pounds_{\phi}B_{ab},\quad \pounds_{\xi}H_{abc}=0=\pounds_{\phi}H_{abc},
\label{b3}
\end{eqnarray}
which gives
\begin{eqnarray}
\pounds_{\chi}B_{ab}=\phi^cB_{c[b}\nabla_{a]}\alpha,\quad \pounds_{\chi}H_{abc}=\phi^dH_{d[bc}\nabla_{a]}\alpha,
\label{b3}
\end{eqnarray}
where the hypersurface orthogonal timelike vector field $\chi^a$ is defined in the previous section. The discrete
symmetry of the spacetime under simultaneous reflections $\xi^a\to-\xi^a$ and $\phi^a\to -\phi^a$ should also
 be obeyed by
any physical matter field. Since the above simultaneous 
reflections imply $\chi_a\to-\chi_a$, we shall set any cross component of $B_{ab}$ along $\chi_{[a}X_{b]}$ or
$\phi_{[a}X_{b]}$, for any $X_a$ orthogonal to both $\chi^a$ and $\phi^a$, to zero. For static spacetimes this statement
will concern only the time-space cross components, as there is in general only time reversal symmetry.

We start with the component $\Psi=(\beta f)^{-1}\chi^a\phi^bB_{ab}$. Contracting Eq.~(\ref{b2}) with $\chi_b\phi_c$,
using Eq.s~(\ref{g5'n}),~(\ref{g9''}) we find
\begin{eqnarray}
\nabla_a\left(\beta f e^a\right)-2fe^a\nabla_a\beta-2\beta e^a\nabla_af-m^2\beta f\Psi=0,
\label{b4}
\end{eqnarray}
where we have defined $ e^a=(\beta f)^{-1}\chi_b\phi_cH^{abc}$. It is clear that $e_a\chi^a=0=e_a\phi^a$. This, along
with the symmetry requirement and the commutativity of the Killing fields give
\begin{eqnarray}
\pounds_{\phi}e_a=0=\pounds_{\phi}\Psi,\quad \pounds_{\chi}e_a=0=\pounds_{\chi}\Psi.
\label{b5}
\end{eqnarray}
Since $\nabla_a\beta$ and $\nabla_a f$ are orthogonal to both $\chi^a$ and $\phi^a$, and so is $e_a$, we shall write
the above equation on the spacelike ({\it n}$-$2)-submanifolds, $\overline{\Sigma}$ using Eq.~(\ref{g8}). We have,
\begin{eqnarray}
\overline{D}_a\left(\beta f e^a\right)=2fe^a\overline{D}_a\beta+2\beta e^a\overline{D}_af+m^2\beta f\Psi+\beta^{-2}\chi^b\chi^a\nabla_a\left(\beta f e_b\right)-f^{-2}\phi^b\phi^a\nabla_a\left(\beta f e_b\right),
\label{b6}
\end{eqnarray}
where $\overline{D}$ is the induced connection on $\overline{\Sigma}$. This equation can be simplified using  
orthogonalities between $e_a$, $\chi_a$ and $\phi_a$, and the Lie derivatives. We find after some calculations a very
simple looking equation,
\begin{eqnarray}
\overline{D}_ae^a-m^2\Psi=0.
\label{b7}
\end{eqnarray}
We also find using Eq.~(\ref{b3}),
\begin{eqnarray}
\beta f e_a=\chi^b\phi^cH_{abc}=\phi^c\left[\nabla_{[a}\left(\chi^bB_{bc]}\right)+\phi^bB_{b[a}\nabla_{c]}\alpha\right].
\label{b8}
\end{eqnarray}
Using $\pounds_{\phi}\alpha=0=\pounds_{\phi}\left(\chi^bB_{bc}\right)$, the above equation further simplifies to 
\begin{eqnarray}
\beta f e_a=\nabla_a\left(\beta f \Psi\right)=\overline{D}_a\left(\beta f \Psi\right).
\label{b9}
\end{eqnarray}
We now multiply Eq.~(\ref{b7}) with $\beta f \Psi $ and use the above equation to get
\begin{eqnarray}
\overline{D}_a\left(\beta f \Psi e^a\right)-\beta f\left[e^ae_a+ m^2\Psi^2\right]=0,
\label{b10}
\end{eqnarray}
which we integrate in the exterior of the black hole horizon. The total divergence can be converted to a surface integral at the boundaries. For $\Lambda\leq0$, the boundaries are black hole horizon $(\beta=0)$ and the spatial infinity,
where we impose sufficient fall-off condition on the matter field, whereas
for the de Sitter case the outer boundary is the de Sitter or cosmological horizon ($\beta=0$). In any case, the surface
integrals go away and we are left with a vanishing integral of positive definites, which shows $\Psi=0=e_a$.  

We shall use this result along with the symmetry arguments to show that the remaining components are vanishing too.
By the requirement of discrete symmetry, there is no other component of $B_{ab}$ which can be directed along $\chi_a$
: $B_{ab}\chi^a=0$,
and hence we have to deal with only purely spatial part of $B_{ab}$. We note that for purely spatial
$B_{ab}$,
\begin{eqnarray}
\chi^aH_{abc}=\pounds_{\chi}B_{ab}=\phi^aB_{ab}\nabla_c\alpha+\phi^aB_{ca}\nabla_b\alpha.
\label{b11}
\end{eqnarray}
By antisymmetry, the quantity $\chi^aH_{abc}$ is purely spacelike. Therefore, since $B_{ab}$ is antisymmetric
the free index in $\phi^aB_{ab}$ must be purely spatial and orthogonal to $\phi_b$. But these components are ruled out
by the discrete symmetry. Thus $\chi^aH_{abc}=0$ and hence $H_{abc}$ is purely spatial.

We now project Eq.~(\ref{b2}) with the help of Eq.~(\ref{g6}) onto $\Sigma$. We find after some algebra,
\begin{eqnarray}
D_a\left(\beta H^{abc}\right)-m^2\beta B^{bc},
\label{b12}
\end{eqnarray}
which we contract with $B_{bc}$ and rewrite as
\begin{eqnarray}
D_a\left(\beta B_{bc} H^{abc}\right)-\beta \left[\frac13H_{abc}H^{abc}+m^2 B^{bc}B_{bc}\right].
\label{b13}
\end{eqnarray}
We integrate this equation as before, and get that all the spatial part of $B_{ab}$ and $H_{abc}$
are vanishing. This completes the no hair proof for massive 2-form fields for stationary axisymmetric spacetimes endowed with
two commuting Killing fields.

We shall now generalize this result for higher form fields in an analogous manner. Let us
consider a free massive 3-form $B_{abc}$ with 4-form field strength $H_{abcd}$, with equation of motion
\begin{eqnarray}
\nabla_aH^{abcd}-m^2B^{bcd}=0,
\label{b18}
\end{eqnarray}
with the totally antisymmetric definition 
\begin{eqnarray}
H_{abcd}=\nabla_aB_{bcd}-\nabla_bB_{cda}+\nabla_cB_{dab}-\nabla_dB_{abc},
\label{b18'}
\end{eqnarray}
and the symmetry conditions
\begin{eqnarray}
\pounds_{\xi}B_{abc}=0=\pounds_{\phi}B_{abc},\quad \pounds_{\xi}H_{abcd}=0=\pounds_{\phi}H_{abcd}.
\label{b18''}
\end{eqnarray}
The requirements from the discrete symmetry applies as well.

Contracting Eq.~(\ref{b18}) with $\chi_c\phi_d$, and using Eq.s~(\ref{g5'n}),~(\ref{g9''}), we get
\begin{eqnarray}
\nabla_a\left(\beta fF^{ab}\right)-2fF^{ab}\nabla_a\beta-2\beta F^{ab}\nabla_a f- m^2\beta f e^b=0,
\label{b19}
\end{eqnarray}
where we have defined $\beta f F^{ab} =H^{abcd}\chi_c\phi_d$, and $\beta f e^a=B^{abc}\chi_b\phi_c$. 
Antisymmetries guarantee that $F_{ab}$ and $e_a$ are orthogonal to both $\chi_a$ and $\phi_a$.
The above is the analogue of Eq.~(\ref{b4}).

Since all the tensors appearing in Eq.~(\ref{b19}) are tangent to $\overline{\Sigma}$, we shall project
it as earlier to get
\begin{eqnarray}
\overline{D}_aF^{ab}-m^2e^b=0.
\label{b20}
\end{eqnarray}
We also have, using Eq.s~(\ref{b18'}), (\ref{b18''}),
\begin{eqnarray}
\beta f F_{ab} =\chi^c\phi^dH_{abcd}=\nabla_a\left(\beta f e_b\right)-\nabla_{b}\left(\beta f e_a\right)=
\overline{D}_{[a}\left(\beta f e_{b]}\right),
\label{b21}
\end{eqnarray}
where in the last equality we have used orthogonalities $e_a\chi^a=0=e_a\phi^a$ as well. Contracting
Eq.~(\ref{b20}) with $\beta f e_b $ we get
\begin{eqnarray}
\overline{D}_a\left(\beta f e_bF^{ab}\right)-\beta f\left[\frac12F_{ab}F^{ab} +m^2e_be^b\right]=0,
\label{b22}
\end{eqnarray}
which we integrate as earlier to get $e_a=0=F_{ab}$ throughout. By the discrete symmetry, the remaining
components of $B_{abc}$ must be purely spatial. Then we can show as earlier that $H_{abcd}$ is purely spatial.
Then we may project Eq.~(\ref{b18}) onto $\Sigma$ to get
\begin{eqnarray}
D_a\left(\beta H^{abcd}\right)-m^2\beta B^{bcd}=0,
\label{b23}
\end{eqnarray}
which we contract with $B_{bcd}$ and integrate by parts to find all the remaining purely spatial components
of $B$ and $H$ to be vanishing.

The process goes on for higher free massive form fields and hence it proves the desired no hair result for
general stationary axisymmetric spacetimes with two commuting Killing fields.

For static spacetimes of arbitrary dimensions, the hypersurface orthogonal timelike vector field $\chi^a$ coincides with the Killing
field $\xi^a$. In this case the various Lie derivatives of the matter fields involve $\xi^a$ only.
Since there is a time reversal symmetry, we set all the space-time cross components to be zero, i.e. 
a massive {\it p}-form $B_{ab\dots}$ is purely spatial. This, along with $\pounds_{\xi}B_{abc\dots}=0$ implies
the ({\it p}$+$1)-form field strength $H=dB$ is also purely spatial. This leads to equation like (\ref{b23}) in
this case from which the no hair result follows. We note that we do not need to use any symmetry
other than $\xi^a$. Hence for static spacetimes, this result is valid irrespective of any spatial symmetry. 

 We were unable to generalize the forgoing results for more than two commuting Killing fields,
as we could not handle the resulting equations to put them in nice forms from which something
meaningful can be extracted. 

\section{Massive spin-$\frac12$ fields}
Let us now come to the massive spin-$\frac12$ case.
We shall assume that the probability density $\Psi^{\dagger}\Psi$ 
associated with a spinor $\Psi$ 
and its derivative is bounded on the horizon (or horizons for de Sitter). We also assume that the norm
of the conserved current $j_a=\overline{\Psi}\gamma_a\Psi$ is bounded there.

Since we are
working with mostly positive metric signature, the anti-commutation for $\gamma$-matrices is
\begin{eqnarray}
[\gamma_a,~\gamma_b]_+=-2g_{ab}\bf{I},
\label{spin1}
\end{eqnarray}
where $g_{ab}$ is the spacetime metric with mostly positive signature. The matrix $\gamma_{0}$
is Hermitian, whereas all the spatial $\gamma$'s are anti-Hermitian.

Let us first consider static black hole spacetimes endowed with a magnetic charge and purely magnetic field.
In this case we shall investigate only the so called zero-energy
solutions.
We work in a gauge in which the gauge field $A_b$ is purely spatial. We note that this 
is in general not possible for stationary axisymmetric spacetimes. The equation of motion is
\begin{eqnarray}
i\gamma^a\widehat{\nabla}_a\Psi-m\Psi=0,\quad
i\widehat{\nabla}_{a}\overline{\Psi}\gamma^a+m\overline{\Psi}=0,
\label{spin2}
\end{eqnarray}
where `$\widehat{\nabla}$' is the gauge-spin covariant derivative : $\widehat{\nabla}_a\Psi
=\nabla_a\Psi-ieA_a\Psi$, and `$\nabla$' is the usual spin covariant derivative. The constant
`$e$' is the charge of the spinor. `Squaring' the first of Eq.s~(\ref{spin2}) we have
\begin{eqnarray}
\widehat{\nabla}^a\widehat{\nabla}_a\Psi+\frac{ie}{2}F_{ab}\gamma^a\gamma^b\Psi-\left(m^2+\frac{R}{4}\right)\Psi=0,
\label{spin3}
\end{eqnarray}
where $F_{ab}$ is the electromagnetic field strength, and $R$ is the Ricci scalar. Taking the Hermitian
conjugate of the above equation, we compute
\begin{eqnarray}
\widehat{\nabla}^a\widehat{\nabla}_a\left(\Psi^{\dagger}\Psi\right)-2\left(\widehat\nabla_a\Psi^{\dagger}\right)
\left(\widehat\nabla^a\Psi\right)
-\Psi^{\dagger}\left(2m^2-ieF_{ij}\gamma^i\gamma^j+\frac{R}{2}\right)\Psi=0,
\label{spin4}
\end{eqnarray}
where $i,~j$ denote purely spatial indices.
We shall now write the above equation in terms of the spacelike derivative operator $D_a$,
using Eq.~(\ref{g6}). By computing $h^{ab}\widehat{\nabla}_a\widehat{\nabla}_b\Psi$ and using
Eq.~(\ref{spin4}), and noting that the hypersurface orthogonal vector field $\chi^a$ coincides
with the Killing field $\xi^a$ in this case, we find after some algebra  
\begin{eqnarray}
D_{a}\left(\beta D^a\left(\Psi^{\dagger}\Psi\right)\right)-2\beta\left(\widehat{D}_a\Psi^{\dagger}\right)
\left(\widehat{D}^a\Psi\right)
-\beta\Psi^{\dagger}\left(2m^2-ieF_{ij}\gamma^i\gamma^j+\frac{R}{2}\right)\Psi-\beta^{-1}
\left\{\xi^a\nabla_a\left(\xi^b\nabla_b\Psi^{\dagger}\right)\Psi+\rm{H.c.}\right\}\nonumber\\=0,
\label{spin5}
\end{eqnarray}
where `H.c.' denotes Hermitian conjugate, and we have used the fact that when a derivative acts on
$\Psi^{\dagger}{\Psi}$, the gauge connection vanishes, and $A_a$ is purely spatial : $A_a\xi^a=0$.
We shall now simplify Eq.~(\ref{spin5}) using the Lie derivative
of spinors~\cite{Godina}. Since in this case we are only investigating zero modes, the spinor $\Psi$ has no
explicit dependence on the parameter along $\xi^a$, which means~\cite{Bhattacharya:2012yt, Godina} 
\begin{eqnarray}
\pounds_{\xi}\Psi=\xi^a\nabla_a\Psi -
\frac14\nabla_{a}\xi_{b}\gamma^a\gamma^b\Psi=0,
\label{spin6}
\end{eqnarray}
which gives the expression for $\xi^a\nabla_a\Psi$. Using this in Eq.~(\ref{spin5}), and using Eq.~(\ref{g18})
(with $\widetilde{\alpha}=0=\alpha$) or Eq.~(\ref{g5'n}) (with $\alpha=0$), and Eq.~(\ref{spin1})
we find after some algebra
\begin{eqnarray}
D_{a}\left(\beta D^a\left(\Psi^{\dagger}\Psi\right)\right)-\beta\left[2\left(\widehat{D}_a\Psi^{\dagger}\right)
\left(\widehat{D}^a\Psi\right)
+\Psi^{\dagger}\left(2m^2-ieF_{ij}\gamma^i\gamma^j+\frac{R}{2}\right)\Psi+\frac{1}{2\beta^2}\left(D_a\beta\right)
\left(D^a\beta\right)\Psi^{\dagger}\Psi\right]=0,\nonumber\\
\label{spin7}
\end{eqnarray}
which we multiply with $\beta$ to write as
\begin{eqnarray}
D_{a}\left(\beta^2 D^a\left(\Psi^{\dagger}\Psi\right)\right)-\beta\left[2\beta\left(\widehat{D}_a\Psi^{\dagger}\right)
\left(\widehat{D}^a\Psi\right)
+\beta\Psi^{\dagger}\left(2m^2-ieF_{ij}\gamma^i\gamma^j+\frac{R}{2}\right)\Psi\right.\nonumber\\
\left.+\frac{1}{2\beta}\left(D_a\beta\right)
\left(D^a\beta\right)\Psi^{\dagger}\Psi+\left(D_a\beta\right)\left(D^a(\Psi^{\dagger}\Psi)\right)\right]=0.
\label{spin8}
\end{eqnarray}
We note that the quantity $\Psi^{\dagger}\Psi$ is not tangent to $\Sigma$, but is the timelike component
of the vector $\overline{\Psi}\gamma^a\Psi$. Also, there can be summation on timelike index in the
spin connection $\omega_{abc}\gamma^b\gamma^c$ associated with $D_a$ (although `$a$' is spacelike).
Clearly, unlike tensors, now there is no natural way to project the entire derivative onto $\Sigma$.
Therefore, the derivative operator `$D$' appearing in the
 above equations should be interpreted as spacelike directional spin-covariant derivative associated with the
full spacetime metric, as $D_a$ acts on quantities not necessarily tangent to $\Sigma$.
Accordingly, when we integrate, we shall use the full invariant volume measure
$[dX]$.
  
We next consider the Killing identity for $\xi_b$,
\begin{eqnarray}
\nabla_a\nabla^a\xi_b=-R_{b}{}^{a}\xi_a,
\label{spin9}
\end{eqnarray}
which we contract with $\xi^b$, use Eq.~(\ref{g18})
(with $\widetilde{\alpha}=0=\alpha$) or Eq.~(\ref{g5'n}) (with $\alpha=0$), to find
\begin{eqnarray}
\nabla_a\nabla^a\beta^2=4\left(\nabla_a\beta\right)\left(\nabla^a\beta\right)+2R_{ab}\xi^a\xi^b,
\label{spin10}
\end{eqnarray}
which we multiply with $\Psi^{\dagger}\Psi$ and rewrite as
\begin{eqnarray}
\nabla_a\left(\Psi^{\dagger}\Psi \nabla^a\beta^2\right)=2\left[2\left(\nabla_a\beta\right)\left(\nabla^a\beta\right)\Psi^{\dagger}\Psi+R_{ab}\xi^a\xi^b\Psi^{\dagger}\Psi+\beta\nabla_{a}\left(\Psi^{\dagger}\Psi\right)(\nabla^a\beta) \right].
\label{spin11}
\end{eqnarray}
We now integrate the above equation using full spacetime volume element $[dX]$. Since the 
1-form $\nabla_a\beta^2$ satisfies Frobenius condition and hence hypersurface orthogonal,
the total divergence can be
converted into surface integrals on the horizon and infinity or on the two horizons (for the de Sitter),
all of which are $\beta^2=\rm{constant}$ hypersurfaces. 
We recall from Section 2 that one of the basis tangent to $\Sigma$ is
$\mu_a=\nabla_a\beta^2$, whose norm vanishes on the horizon(s) as ${\cal{O}}(\beta^2)$,
Eq.s~(\ref{g21}), (\ref{g22}). Now the surface integral at horizon(s)
looks like $\int_{{\cal{H}}}\Psi^{\dagger}\Psi \nabla_a\beta^2ds^{a}$, where $ds^{a}$ in the 
(n-1)-dimensional volume element, with the unit normal directing along $\mu^a$. Since the volume element on
the horizon contains a $\beta$, and we have assumed the quantity $\Psi^{\dagger}\Psi$ is bounded on
the horizons, the above surface integral is bounded there, and it contains $\mu^a\nabla_a\beta^2$.
But this is vanishing on the horizon(s), so the integral on horizon(s) vanish. For $\Lambda\leq0$,
we impose sufficiently rapid fall-off on $\Psi$ at infinity, so that it vanishes there too,
leaving us with the vanishing volume integral of the right hand side of Eq.~(\ref{spin11}).
We now combine this with the integral of Eq.~(\ref{spin8}), recalling $\nabla_a\beta=D_a\beta$,
we find after some rearrangement
\begin{eqnarray}
\int[dX]\left[2\left(\beta\widehat{D}_a\Psi+\Psi D_a\beta\right)^{\dagger} \left(\beta\widehat{D}^a\Psi+\Psi D^a\beta\right)+2\beta^2\Psi^{\dagger}\left(m^2-\frac{ie}{2}F_{ij}\gamma^i\gamma^j\right)\Psi\right.\nonumber\\
\left.
+\left(R_{ab}-\frac12 R g_{ab}\right)\xi^a\xi^b \Psi^{\dagger}\Psi+\frac12 \left(D_a\beta\right)
\left(D^a\beta\right)\Psi^{\dagger}\Psi  \right]=0,
\label{spin12}
\end{eqnarray}
and using Einstein's equations $R_{ab}-\frac12 Rg_{ab}+\Lambda g_{ab}=T_{ab}$, we have
\begin{eqnarray}
\int[dX]\left[2\left(\beta\widehat{D}_a\Psi+\Psi D_a\beta\right)^{\dagger} \left(\beta\widehat{D}^a\Psi+\Psi D^a\beta\right)+2\beta^2\Psi^{\dagger}\left(m^2+\frac{\Lambda}{2}-\frac{ie}{2}F_{ij}\gamma^i\gamma^j\right)\Psi\right.\nonumber\\
\left.
+T_{cd}\xi^c\xi^d \Psi^{\dagger}\Psi+\frac12 \left(D_a\beta\right)
\left(D^a\beta\right)\Psi^{\dagger}\Psi  \right]=0.
\label{spin13}
\end{eqnarray}
Since the index `$a$' is spatial in the above equation, the first and the last terms are positive definite,
whereas the third term is positive definite since the Maxwell field obeys weak and null energy conditions
and fermion's backreaction has been ignored.
Let us now examine the second term which contains the so called (Hermitian) anomalous correction to the fermion mass.  
Since the black hole is the source of the magnetic charge, the magnetic field should decrease with distance 
from the black hole horizon. If $Q$ is the magnetic charge of the black hole, we must have in our units
the quantity $eQ$ to obey some certain smallness conditions, otherwise since $A_a$ contains a $Q$,
 the term corresponding to 
$e\overline{\Psi}\gamma^a\Psi A_a$ would backreact into the energy-momentum tensor. Also as we have
discussed at the end of Section 2, for backreactionless massive fields the Compton wavelength
($\sim m^{-1}$) of the field is small compared to the black hole horizon size. Since the anomalous term
has dimensions $\sim\rm{length}^{-2}$, it is clear that the mass term should dominate it everywhere
outside the black hole horizon. For $\Lambda<0$, the mass term would dominate the $\Lambda$ term too,
since the AdS length scale should obviously be larger than the black hole length scale. Putting these
all in together, we find that a magnetic static black hole spacetime of
arbitrary dimension cannot support fermion zero modes in its exterior, i.e. $\Psi=0$,
provided we can ignore fermion's beackreation. 

An immediate corollary of the above result for neutral fermions
is obtained by setting $e=0$, for black holes with electric and/or magnetic 
(non-)Abelian charge(s). We note that unless we used the Killing identity for $\xi^b$,
we could not have obtained this conclusion.
 
We shall now consider electrically charged static black holes, assuming 
there is only electric field. From now on we do not need to confine to the zero modes only.
Following~\cite{Bhattacharya:2012yt}, we define a 2-form $S_{ab}$ from the conserved current 1-form
$j_a=\overline{\Psi}\gamma_a\Psi$,
\begin{eqnarray}
S_{ab}=\nabla_aj_b-\nabla_bj_a,
\label{spin14}
\end{eqnarray}
so that 
\begin{eqnarray}
\nabla^aS_{ab}=\nabla_a\nabla^aj_b-R_{b}{}^{a}j_a,
\label{spin15}
\end{eqnarray}
using $\nabla_aj^a=0$. After using Einstein's equations in {\it n}-dimensions, each component $\lambda$
of the above equation becomes
\begin{eqnarray}
\nabla^aS_{a\lambda}=\nabla_a\nabla^aj_\lambda-\left[T_{a\lambda}j^a-\frac{T}{n-2}j_\lambda+\frac{2\Lambda}{n-2}j_\lambda\right],
\label{spin16}
\end{eqnarray}
where $T_{ab}$ does not contain the fermion contribution as earlier.
We integrate the above equation between the black hole horizon and infinity (or the cosmological horizon for de Sitter).
The total divergences can be converted into surface integrals and on the horizon(s) have the form
\begin{eqnarray}
\int_{ {\cal{H}} }S_{a\lambda}ds^a-\int_{{\cal{H}}}\left(\nabla_aj_{\lambda}\right)ds^a,
\label{spin17}
\end{eqnarray}
where as before the direction `$a$' corresponds to a unit vector along the basis $\mu_a=\nabla_a\beta^2$,
which becomes null and coincides with $\xi_a$ on ${\cal{H}}$. For $\Lambda\leq0$, imposing 
suitable fall-off at infinity makes the surface integral vanishing there.
We have by the symmetry requirement,
\begin{eqnarray}
\pounds_{\xi}\left(\overline{\Psi}\gamma_a\Psi\right)=0,
\label{spin18}
\end{eqnarray}
which gives after using Eq.~(\ref{g5'n}) (with $\alpha=0$ for static case)
\begin{eqnarray}
\xi^b\nabla_bj_a=\Psi^{\dagger}\Psi\nabla_a\beta-\xi_a\frac{1}{2\beta^2}j^b\nabla_b\beta^2,
\label{spin19}
\end{eqnarray}
and set $a=0$. The first term goes away 
$({\rm since}~\xi^a\nabla_a\beta=0,~{\rm everywhere})$, and we are left with the second term only,
and $\xi_0$ can be taken as $-\beta^2$. 

Now we can evaluate the directional covariant derivative
in the second of Eq.~(\ref{spin17})
on the horizon(s), where $\mu_a=\nabla_a\beta^2$ coincides with $\xi_a$ and its norm
vanishes as ${\cal{O}}(\beta^2)$ there. We recall our assumption that both $\Psi^{\dagger}\Psi$
and $j_aj^a$ are bounded on the horizon(s), which implies $\beta^{-1}j^a\nabla_a\beta^2$
is also bounded on the horizon(s).
Thus it is clear that $\xi^b\nabla_bj_{\lambda}\vert_{\lambda=0}$ is vanishing as 
at least ${\cal{O}}(\beta)$ on the horizon(s).
Thus the second integral in Eq.~(\ref{spin17}) is vanishing. Another way to see this is to
integrate the conservation equation $\nabla_aj^a=0$, and convert it to surface integrals on
the horizon and infinity (or on horizons for de Sitter), and since the surface 
integrand is $j_a\mu^a$, the result follows from comparison with Eq.~(\ref{spin19}) with $a=0$.

On the other hand, since $S_{ab}$ is antisymmetric in its indices, and $\mu^a$ coincides with
$\xi^a$ on the horizon(s), we have $S_{0b\mu^b}=0$ there. 

Putting these all in together, we see that setting $\lambda=0$ in Eq.~(\ref{spin16}) means  
\begin{eqnarray}
\int[dX]\left[T_{a0}j^a+\frac{\beta T}{n-2}\Psi^{\dagger}\Psi-\frac{2\beta\Lambda}{n-2}\Psi^{\dagger}\Psi\right]=0.
\label{spin20}
\end{eqnarray}
We next note that $j_a$ can never be spacelike. Although this is obvious intuitively, but can be proven
as the following. If possible we assume that $j_a$ is spacelike in some region of the spacetime. We erect
a local Lorentz frame at some point $P$ in this region and rotate it to one of the spatial axis of this frame
to coincide with $j_a$. But this will mean the `time' component of $j_a$ to be vanishing identically,
which means $\Psi=0$. This is clearly a contradiction, and hence $j_a$ must be non-spacelike 
and future directed. In particular,
since we are dealing with massive fields, it must be timelike. Then the energy condition
discussed at the beginning of Section 2 guarantees that the first term in Eq.~(\ref{spin20}) is positive.
The third term is positive or zero if $\Lambda\leq 0$, whereas for Maxwell field 
$T=\left(1-\frac{n}{4}\right)F_{ab}F^{ab}\geq 0$ if there are only electric fields. This shows that
for $\Lambda\leq 0$, there can be no Dirac hair for static electrically charged spacetimes endowed only with
electric fields, provided we can ignore backreaction of spinors and Eq.~(\ref{spin18}) is satisfied. This
of course include real frequency solutions, when $\xi^a$ is a coordinate vector field.  
 
What happens if we are working in four spacetime dimensions with $\Lambda=0$? Eq.~(\ref{spin20})
then only contains the first term, which is positive definite. Hence we must have $T_{a0}j^a=0$
throughout. We next decompose $T_{a0}$ along $j_a$ and orthogonal to it. Since $F_{ab}$
is non-vanishing, we must have $j_aj^a=0$ throughout. But $j_a$ is future
directed timelike, so this is a contradiction. Therefore we must have $j_a=0$, which means
$\Psi=0$ throughout.

We were unable to find an analogous proof for $\Lambda>0$. Perhaps
there are some additional conditions or identities which should be used (as we used Killing identity
in the previous part), but we were unable to find any. 

We shall conclude this Section by noting the following for $\Lambda$-vacuum (positive or negative)
stationary axisymmetric spacetimes in arbitrary dimensions with arbitrary number of commuting Killing
fields. Let us first consider a stationary axisymmetric spacetime with
three commuting Killing fields as discussed in Section 2. The symmetry requirement in this case becomes
\begin{eqnarray}
\pounds_{\xi}j_a=\pounds_{\phi}j_a=\pounds_{\phi^1}j_a=0.
\label{spin21}
\end{eqnarray}
The hypersurface orthogonal vector field $\chi_a$ is given by Eq.~(\ref{g10}). Then
 using Eq.s~(\ref{spin21}), (\ref{g18}) we find in place of Eq.~(\ref{spin19})
%
\begin{eqnarray}
\chi^a\nabla_aj_b=\Psi^{\dagger}\Psi\nabla_b\beta+\frac{\chi_b}{2\beta^2}j^a\nabla_a\beta^2
-\frac12 j^a\left[\widetilde{\phi}_{(a}\nabla_{b)}\widetilde{\alpha}+\phi_{(a}\left(\nabla_{b)}\alpha+
\widetilde{\alpha}\nabla_{b)}\lambda\right)\right]+\nonumber\\\left[j_a\phi^a\left(\nabla_b\alpha+\widetilde{\alpha}
\nabla_b\lambda\right)+\widetilde{\phi}^aj_a\nabla_b\widetilde\alpha  \right].
\label{spin22}
\end{eqnarray}
Using Eq.s~({\ref{g13'}}) and $\chi_a\phi^a=0=\chi_a \widetilde{\phi}^a$, we have
\begin{eqnarray}
\chi^a\nabla_aj_\lambda\vert_{\lambda=0}=\frac{\chi_0}{2\beta^2}j^a\nabla_a\beta^2,
\label{spin23}
\end{eqnarray}
which is formally the same as the static case. Consequently by our choice of basis we arrive at 
Eq.~(\ref{spin20}) with only the third term. This guarantees $\Psi=0$ throughout. Since
it is clear that for an arbitrary stationary axisymmetric spacetime with commuting
but non-orthogonal Killing fields $\{\xi,~\phi,~\phi^1,~\phi^2,~\dots\}$ with integral spacelike
submanifolds orthogonal to these Killing fields, the hypersurface orthogonal timelike 
vector field $\chi_a$ can be constructed from their linear combinations and the Killing horizon(s)
can be specified, we conclude that for stationary axisymmetric (anti)-de Sitter spacetimes
falling into the category we discussed in Section 2, there can be no backreactionless Dirac hair with
real phases. 

\section{Discussions}
It is time to summarize the various results we obtained in this paper. Using the necessary
geometrical set up and assumptions described in Section 2, we investigated no hair properties 
of general stationary axisymmetric and static black hole spacetimes. In Section 3, we demonstrated
the no hair proof for massive 2- and higher forms. In the next Section we discussed the case of massive
spin-$\frac12$ fields.

Apart from symmetries, energy conditions, and regularities,
we have not used any particular functional form of the metric or matter fields,
 the reason is the so far not very well
understood uniqueness nature of black hole spacetimes in higher dimensions, with or without $\Lambda$.
We did not have to perform any
complicated variable separations, which even in four spacetime dimensions is a formidable
task. We note that our
results are also valid even if we are not dealing with an exact solution of Einstein's equations.
As long as the spacetime falls into the category described in Section 2, our calculations apply.
An example of this would be the axisymmetric static spacetime constructed in~\cite{Chandrasekhar:1985kt}.
An exact solution of stationary axisymmetric black hole spacetime with three commuting Killing vector
fields can be found in~\cite{Chong:2005hr}. Our analysis is valid for spacetimes with Killing horizons like
 black string spacetimes
(see e.g.~\cite{Kurita:2008mj, Horowitz:2002ym}), for black holes
with toroidal topology~\cite{Rinaldi:2002tc}, and as well as for the black rings~\cite{Emparan:2001wn}.
Our result is also valid for multi-black hole spacetimes described in e.g.~\cite{Chandrasekhar:1985kt}, which
is not spherically symmetric, and not necessarily be axisymmetric as well,
or other multi horizon black hole spacetimes in higher dimensions (see~\cite{Emparan:2008eg} for a vast review and list
of references),
only we have to replace the inner boundary integral with the sum of integrals on all the black hole 
horizons, since we have considered the horizon(s) in a purely geometric way as $\beta^2=0$ 
null hypersurface(s).    

It remains as an interesting task to further generalize the massive spin-$\frac12$ result for 
arbitrary stationary axisymmetric spacetimes carrying electric or magnetic charge.


\section*{Acknowledgment}
I thank Amitabha Lahiri for useful discussions.

\vskip 1cm


\begin{thebibliography}{99}


\bibitem{Chrusciel:1994sn}
  P.~T.~Chrusciel,
  Contemp.\ Math.\  {\bf 170}, 23 (1994).



\bibitem{Heusler:1998ua}
  M.~Heusler,
  Living Rev.\ Rel.\  {\bf 1}, 6 (1998).


\bibitem{Heusler}
  M.~Heusler,
  ``Black Hole Uniqueness Theorems,''
{\it  Cambridge Univ. Pr. ( 1996)}.


\bibitem{Bekenstein:1998aw}
  J.~D.~Bekenstein,
arXiv:gr-qc/9808028.


\bibitem{Bekenstein:1971hc}
  J.~D.~Bekenstein,
  Phys.\ Rev.\ D {\bf 5}, 1239 (1972).


\bibitem{Bekenstein:1972ky}
  J.~D.~Bekenstein,
  Phys.\ Rev.\  D {\bf 5}, 2403 (1972).

\bibitem{Bhattacharya:2007ap} 
  S.~Bhattacharya and A.~Lahiri,
  Phys.\ Rev.\ Lett.\  {\bf 99}, 201101 (2007).

\bibitem{Bhattacharya:2011dq} 
  S.~Bhattacharya and A.~Lahiri,
  Phys.\ Rev.\ D {\bf 83}, 124017 (2011).

\bibitem{Bhattacharya:2012yt} 
  S.~Bhattacharya and A.~Lahiri,
  Phys.\ Rev.\ D {\bf 86}, 084038 (2012).


\bibitem{Anabalon:2012ta} 
  A.~Anabalon,
  JHEP {\bf 1206}, 127 (2012).


\bibitem{Anabalon:2009qt} 
  A.~Anabalon and H.~Maeda,
  Phys.\ Rev.\ D {\bf 81}, 041501 (2010).


\bibitem{Sen:1998bj} 
  S.~Sen and N.~Banerjee,
 Pramana, {\bf 56}, 487 (2001) [arxiv:gr-qc/9809064].

\bibitem{Anabalon:2012tu} 
  A.~Anabalon and A.~Cisterna,
  Phys.\ Rev.\ D {\bf 85}, 084035 (2012).


\bibitem{Anabalon:2012sn} 
  A.~Anabalon, F.~Canfora, A.~Giacomini and J.~Oliva,
  JHEP {\bf 1206}, 010 (2012).

\bibitem{Anabalon:2012ih} 
  A.~Anabalon and J.~Oliva,
  Phys.\ Rev.\ D {\bf 86}, 107501 (2012).

\bibitem{Acena:2012mr} 
  A.~Acena, A.~Anabalon and D.~Astefanesei,
  Phys.\ Rev.\ D {\bf 87}, 124033 (2013). 



\bibitem{Johannsen:2012ng} 
  T.~Johannsen and D.~Psaltis,
 Astrophys.\ J.\ {\bf 773}, 57 (2013). 



\bibitem{Rodriguez:2011aa} 
  C.~L.~Rodriguez, I.~Mandel and J.~R.~Gair,

  Phys.\ Rev.\ D {\bf 85}, 062002 (2012). 


\bibitem{Johannsen:2011dh} 
  T.~Johannsen and D.~Psaltis,
  Phys.\ Rev.\ D {\bf 83}, 124015 (2011).



\bibitem{Shiromizu:2011he} 
  T.~Shiromizu, S.~Ohashi and K.~Tanabe,
  Phys.\ Rev.\ D {\bf 83}, 084016 (2011).



\bibitem{Allen:1989kc} 
  T.~J.~Allen, M.~J.~Bowick and A.~Lahiri,
  Phys.\ Lett.\ B {\bf 237}, 47 (1990).







\bibitem{Moderski:2008nq}
  R.~Moderski and M.~Rogatko,
  Phys.\ Rev.\  D {\bf 77}, 124007 (2008).


\bibitem{Gibbons:2008gg}
  G.~W.~Gibbons, M.~Rogatko and A.~Szyplowska,
  Phys.\ Rev.\  D {\bf 77}, 064024 (2008).


\bibitem{Gibbons:2008rs}
  G.~W.~Gibbons and M.~Rogatko,
  Phys.\ Rev.\  D {\bf 77}, 044034 (2008).


\bibitem{Nakonieczny:2012df} 
  L.~Nakonieczny and M.~Rogatko,
  Phys.\ Rev.\ D {\bf 85}, 124050 (2012).


\bibitem{Gozdz:2010wv} 
  M.~Gozdz, L.~Nakonieczny and M.~Rogatko,
  Phys.\ Rev.\ D {\bf 81}, 104027 (2010).

\bibitem{Price:1971gc}
  R.~H.~Price,
  Phys.\ Rev.\ D {\bf 5}, 2439 (1972).

\bibitem{Chambers:1994sz}
  C.~M.~Chambers and I.~G.~Moss,
  Phys.\ Rev.\ Lett.\  {\bf 73}, 617 (1994).


\bibitem{Chandrasekhar:1985kt} 
  S.~Chandrasekhar,
  ``The mathematical theory of black holes,''
  OXFORD, UK: CLARENDON (1985).


\bibitem{Finster:1998ju} 
  F.~Finster, J.~Smoller and S.~-T.~Yau,
Commun.\ Math.\ Phys.\ {\bf 205}, 249 (1999).

\bibitem{Finster:1998ak} 
  F.~Finster, J.~Smoller and S.~-T.~Yau,
  J.\ Math.\ Phys.\  {\bf 41}, 2173 (2000).


\bibitem{Finster:2001vn} 
  F.~Finster, N.~Kamran, J.~Smoller and S.~-T.~Yau,
  Commun.\ Math.\ Phys.\  {\bf 230}, 201 (2002).



\bibitem{Finster:2000jz} 
  F.~Finster, N.~Kamran, J.~Smoller and S.~-T.~Yau,
  Adv.\ Theor.\ Math.\ Phys.\  {\bf 7}, 25 (2003).


\bibitem{Finster:2005dg} 
  F.~Finster, N.~Kamran, J.~Smoller and S.~-T.~Yau,
  Commun.\ Math.\ Phys.\  {\bf 264}, 465 (2006).


\bibitem{Finster:1999tt} 
  F.~Finster, J.~A.~Smoller and S.~-T.~Yau,
Meth.\ Appl.\ Anal.\ {\bf 8}, 623 (2001).


\bibitem{Finster:1999ry} 
  F.~Finster, N.~Kamran, J.~Smoller and S.~-T.~Yau,
  Commun.\ Pure Appl.\ Math.\  {\bf 53}, 902 (2000).


\bibitem{Finster:2000vy} 
  F.~Finster, J.~Smoller and S.~-T.~Yau,
  Adv.\ Theor.\ Math.\ Phys.\  {\bf 4}, 1231 (2002).

\bibitem{Finster:2000ps} 
  F.~Finster, J.~Smoller and S.~-T.~Yau,
  Nucl.\ Phys.\ B {\bf 584}, 387 (2000).


\bibitem{Belgiorno:2008mx} 
  F.~Belgiorno and S.~L.~Cacciatori,
  Phys.\ Rev.\ D {\bf 79}, 124024 (2009).


\bibitem{Meierovich}
B.~E.~Meierovich,
Phys.\ Rev.\ D {\bf 87}, 103510 (2013).


\bibitem{Prokopec:2006kr} 
  T.~Prokopec and W.~Valkenburg,
  astro-ph/0606315.






\bibitem{Ehlers:2003tv} 
  J.~Ehlers and R.~P.~Geroch,
  Annals Phys.\  {\bf 309}, 232 (2004).

\bibitem{Bhattacharya:2013trx} 
  S.~Bhattacharya, Ph.D. thesis, Jadavpur University, 2013,
  arXiv:1302.1399.

\bibitem{Wald:1984rg}
  R.~M.~Wald,
  ``General Relativity,''
{\it  Chicago, Usa: Univ. Pr. (1984)}.

\bibitem{weinberg}
  S.~Weinberg,
  ``Gravitation and Cosmology,''
{\it  John Wiley and Sons, New York (1972)}.


\bibitem{Penrose:1985jw}
  R.~Penrose and W.~Rindler,
  ``Spinors And Space-Time. 1. Two Spinor Calculus And Relativistic
  Fields,'' 
{\it  Cambridge, Uk: Univ. Pr. (1984)}. 


\bibitem{perry} 
R.~Myers and M.~J.~Perry, Annals Phys.\ {\bf 172} 304 (1986).

\bibitem{Godina}
  M.~Godina and P.~Matteucci,
Int.\ J.\ Geom.\ Methods\ Mod.\ Phys., {\bf 2}, 159 (2005)
[arXiv:math/0504366].  

\bibitem{Chong:2005hr}
  Z.~W.~S.~Chong, M.~Cvetic, H.~Lu and C.~N.~Pope,
  Phys.\ Rev.\ Lett.\  {\bf 95}, 161301 (2005).

\bibitem{Kurita:2008mj}
  Y.~Kurita and H.~Ishihara,
  Class.\ Quant.\ Grav.\  {\bf 25}, 085006 (2008).


\bibitem{Horowitz:2002ym}
  G.~T.~Horowitz and K.~Maeda,
  Phys.\ Rev.\  D {\bf 65}, 104028 (2002).

\bibitem{Rinaldi:2002tc}
  M.~Rinaldi,
  Phys.\ Lett.\  B {\bf 547}, 95 (2002).

\bibitem{Emparan:2001wn} 
  R.~Emparan and H.~S.~Reall,
  Phys.\ Rev.\ Lett.\  {\bf 88}, 101101 (2002).


\bibitem{Emparan:2008eg} 
  R.~Emparan and H.~S.~Reall,
  Living Rev.\ Rel.\  {\bf 11}, 6 (2008).


\end{thebibliography}
\end{document}